\documentclass[a4paper,11pt]{article}
\pdfoutput=1 % if your are submitting a pdflatex (i.e. if you have
\usepackage{jinstpub} % for details on the use of the package, please
\title{\boldmath Temperature scaling of reverse current generated in proton irradiated silicon bulk}

\author[1]{F.~Wizemann,\note{Corresponding author.}}
\author{A.~Gisen,}
\author{K.~Kr\"oninger,}
\author{and J.~Weingarten}
\affiliation{Technische Universit\"at Dortmund, Experimentelle Physik IV, 44221 Dortmund, Germany}

\emailAdd{felix.wizemann@tu-dortmund.de}

\abstract{
The value of the scaling parameter \eeff of the temperature dependence for current generated in silicon bulk is investigated for highly irradiated devices.

Measurements of devices irradiated to fluences above $1 \times 10^{15} \, \neqpcm$ have shown a different temperature scaling behaviour than devices irradiated to lower fluences.
This paper presents the determination of the parameter \eeff for diodes irradiated with protons up to fluences of \SI{$3 \times 10^{15}$}{\neqpcm} in the bias range from $0\,$V to $1000\,$V at temperatures from $-36\celsius$ to $0\celsius$ at different stages of annealing. 
It is shown that \eeff for highly irradiated devices depends on the applied electric field: below depletion voltage, \eeff is observed to have a lower value than above depletion voltage.
}


\usepackage{subcaption}
\graphicspath{{figures/}} % Directory in which figures are stored

\newcommand{\eeff}{\ensuremath{E_{\text{eff}}\ }}
\newcommand{\eeffnospace}{\ensuremath{E_{\text{eff}}}}

\newcommand{\neqpcm}{\ensuremath{n_{\text{eq}} \text{cm}^{-2}}}

\newcommand{\micro}{\textmu}

\newcommand{\meter}{m}

\newcommand{\celsius}{\ensuremath{\,^{\circ}\text{C}}}

\newcommand{\SI}[2]{{#1}\,{#2}}

\begin{document}
\maketitle
\flushbottom

\section{Introduction}
\label{sec:intro}
Current and future tracking detectors in high energy particle physics rely on a variety of silicon-based sensors which are exposed to high radiation levels during operation. The resulting radiation damage has significant influence on the sensor characteristics, including an increased leakage current and depletion voltage.
These effects result in an increased power dissipation which presents a significant challenge for the design of cooling systems.
It is therefore necessary to predict the leakage current of silicon devices after irradiation for different operational bias voltages and temperatures.

The study presented here focuses on the temperature-dependent change of the leakage current in irradiated silicon devices. The effective energy $\eeffnospace\,$\cite{Chili13} is used as a scaling parameter in the parametrization of the temperature dependence.
A decrease of \eeff at high fluences has been observed \cite{Chili13,Wiehe,Wonsak} which is not explained by the model established in ref. \cite{Chili13}.
This study investigates the influence of bias voltage and annealing on this effect. 

Section \ref{sec:theory} summarizes the theoretical background. The experimental method is described in section \ref{sec:meth}. The results are presented in section \ref{sec:results}, discussed in section \ref{sec:disc} and summarized in section~\ref{sec:summary}.

\section{Theoretical background}
\label{sec:theory}
Semiconductor detectors in high energy physics use pn-junctions under reverse biasing to generate a depleted volume to detect energy deposited in the detector material by traversing charged particles.

The leakage current generated in the depleted volume scales with temperature according to
\begin{equation}
    \label{eq:dep}
   I(T) \propto T^2 \exp\left(- \frac{\eeffnospace}{2 k T}\right)
\end{equation}
with leakage current $I$, temperature $T$, Boltzmann constant $k$ and effective energy \eeff \cite{Chili13}.
According to the model presented in ref. \cite{Chili13}, \eeff is expected to be around $1.21\,$eV for current generated in mid level gaps using values for the band gap energy $E_{g}$ in the applicable temperature range presented in ref. \cite{Green}. 

\section{Methodology}
\label{sec:meth}
\subsection{Samples}
The samples used are n-bulk diodes with a thickness of \SI{250}{\micro\meter}. Their central p$^+$ implant has an area of $9\,$mm$^2$ and is surrounded by 16 guard rings. All samples were irradiated with protons at the IRRAD facility\footnote{The CERN Proton Irradiation Facility (IRRAD) uses $24\,$GeV/c protons from the Proton Synchroton.}.
Due to a failure of the cooling system during irradiation, samples P1, P3 and P4 were irradiated without cooling and experienced an unknown amount of annealing. 
The annealing times stated in this paper only account for intentional, monitored annealing afterwards.
Their respective fluences and annealing times are listed in table \ref{tab:samples}. 
The diodes were used in previous studies (in case of sample P1 with intentional, monitored annealing). 

\begin{table}[h]
\centering
\caption{\label{tab:samples} Investigated samples with their respective fluences and annealing times at 60\celsius.}
\smallskip
\begin{tabular}{|c|c|c|}
\hline
Sample & Fluence & Annealing range \\
&[$10^{15}$\neqpcm]& [min]\\
\hline
P1 & 0.6& 1170\\
P3 & 0.7& 0 \text{to} 1800\\
P4 & 3& 0 \text{to} 1170\\
P5 & 1& 0 \text{to} 1800\\
\hline
\end{tabular}
\end{table}

\begin{figure}[ht]
    \begin{subfigure}{0.49\textwidth}
        
        \centering
        \includegraphics[width=1.0\textwidth]{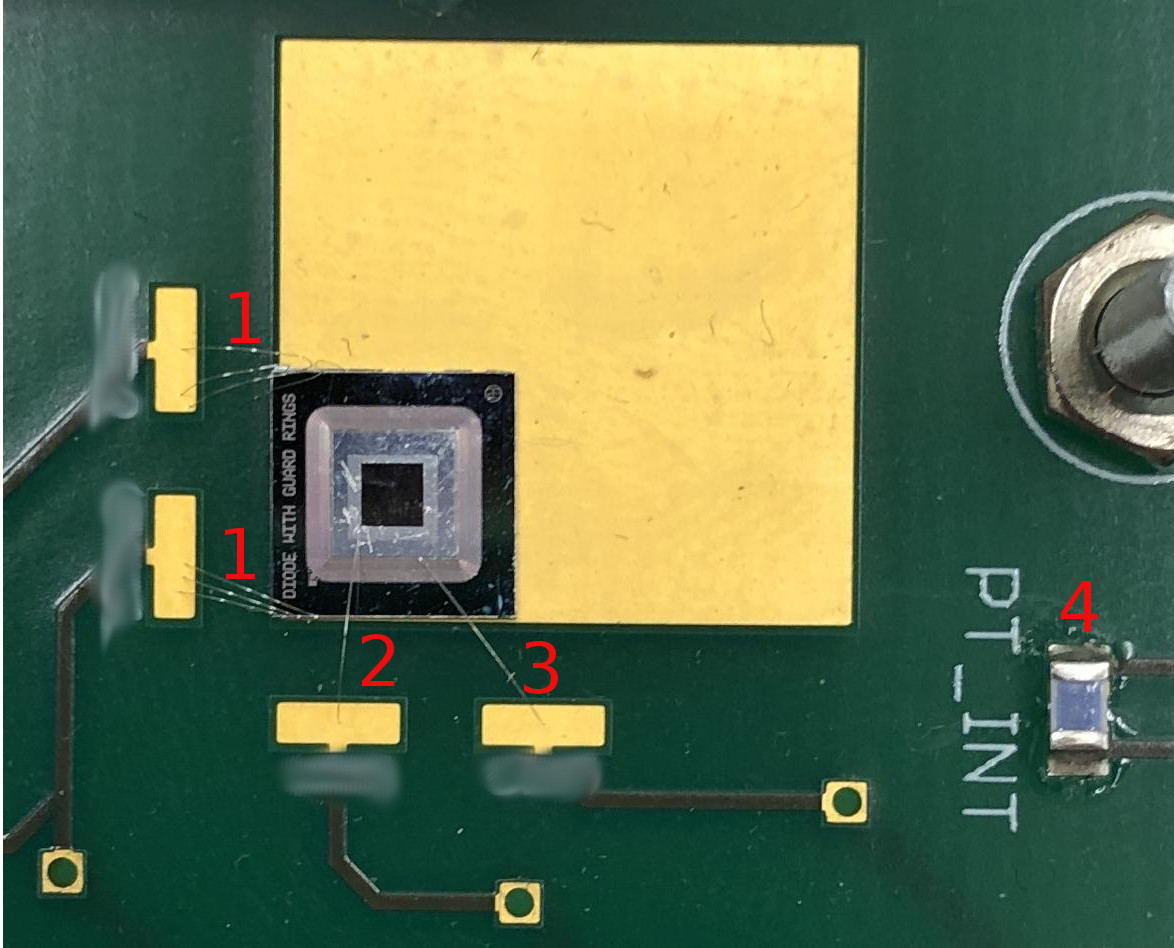}
        \caption{\label{fig:PCB}}
    \end{subfigure}
    \hfill
    \begin{subfigure}{0.49\textwidth}
        
        \centering
        \includegraphics[width=.8\textwidth]{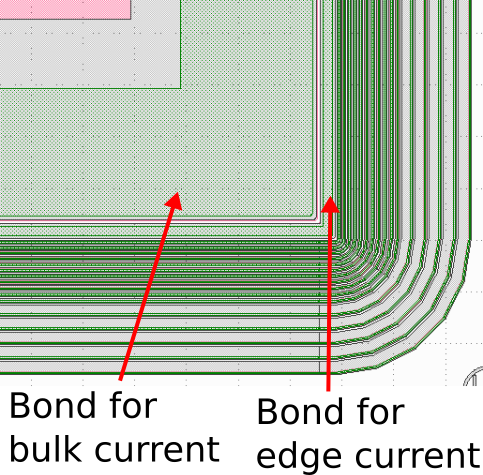}
        \caption{\label{fig:Layout}}
    \end{subfigure}
    \caption{a) Sampe glued on a PCB, including bonds to connect the n-side (1), central p-implant (2), and the innermost guard-ring (3). The thermistor (4) is visible in the lower right corner. b)~Schematic layout of the sample diodes. The innermost ring around the central p-implant is used to separate the edge current from the bulk current.}
\end{figure}

The diodes are glued to a PCB and contacted with wirebonds as shown in figure \ref{fig:PCB}. 
The remaining metal of the dicing street of the diode connects via the dicing edge to the n$^+$ implant on the backside.
Therefore, wirebonds onto the metal of the dicing street are used to apply bias voltage to the n-side of the diode (marked with ``1'').
For the ground contact, the central implant (marked with ``2'') and the innermost guard ring (marked with ``3'') are contacted separately, which can be seen in figure \ref{fig:Layout}.
This serves to measure the bulk current $I_{\text{b}}$ through the central p$^+$ implant to determine \eeff as well as the total current $I_{\text{t}}$ (including the bulk and the surface current) to determine the power dissipation.

\subsection{Measurements}

\begin{figure}
        \centering
        \includegraphics[width=.7\textwidth]{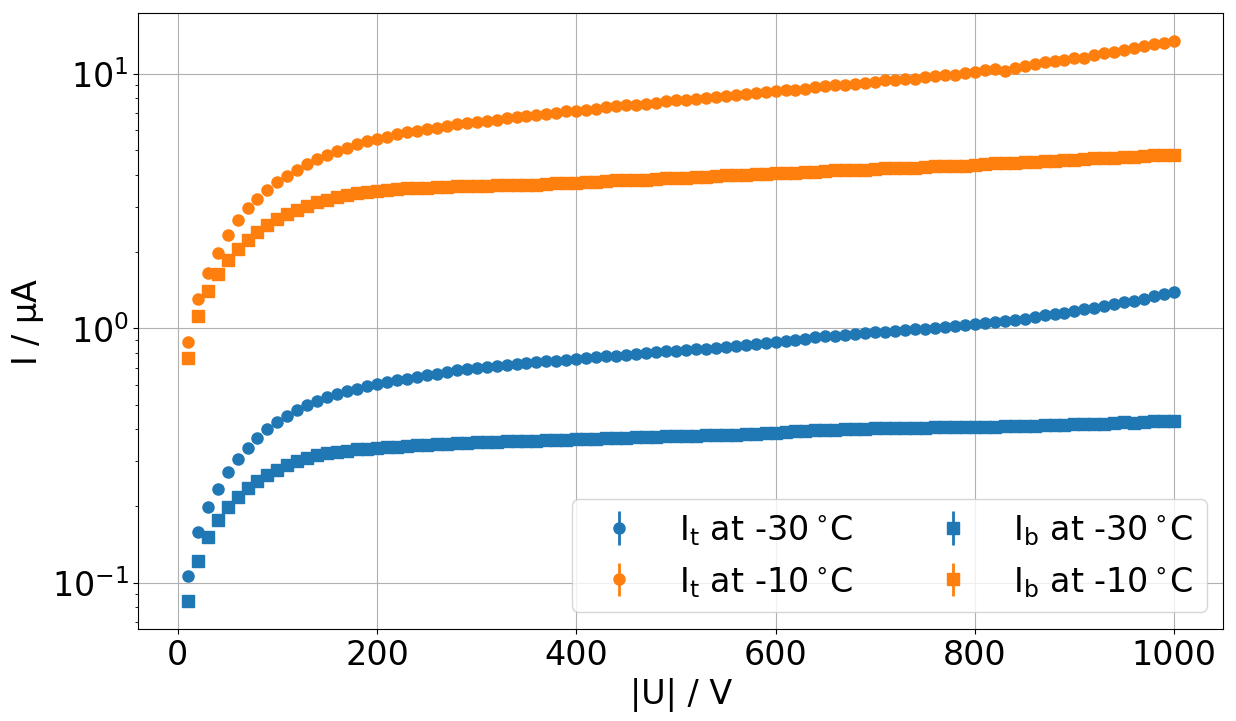}
        \caption{\label{fig:bulkcmp}Comparison of $I_{\text{t}}$ and $I_{\text{b}}$ versus bias voltage for diode P3 at two different temperatures.}
\end{figure}

Figure \ref{fig:bulkcmp} shows the difference between $I_{\text{t}}$ and $I_{\text{b}}$ for the sample P3.
$I_{\text{b}}$ exhibits the expected behaviour for bulk current with an increase in leakage current until depletion is reached and a plateau at higher voltages. A steady increase in current at higher voltages can be seen for $I_{\text{t}}$ implying significant contributions to the total current by sources other than the bulk current.

The determination of \eeff is based on measurements of current-voltage (I-V) characteristics up to $1000\,$V at temperatures from $-36\celsius$ to $0\celsius$ in steps of $2\celsius$.
These measurements are performed inside a climate chamber flushed with dry air. 
The temperature of the sample is monitored during the measurements with a \textit{Pt-1000} thermal resistor on the PCB close to the diode (see figure \ref{fig:PCB}).
Annealing of the diodes is also done in the climate chamber setup.

To regulate temperature, the climate chamber operates in cycles which results in periodically fluctuating temperatures.
Long term measurements at constant temperature and voltage were performed which revealed the periodic changes in the measured leakage current due to the temperature fluctuations.
After correcting for the temperature changes, measurements of the current show a standard deviation of $0.1\%$. This is due to a phase shift between the changes in the temperature of the thermal resistor and the leakage current of the diode. 
The period of these cycles is far longer than the time spent at a single voltage during the measurement of the I-V characteristics, therefore the measured leakage current cannot be corrected for this effect.

\subsection{Power limit}

\begin{figure}
        \centering
        \includegraphics[width=.7\textwidth]{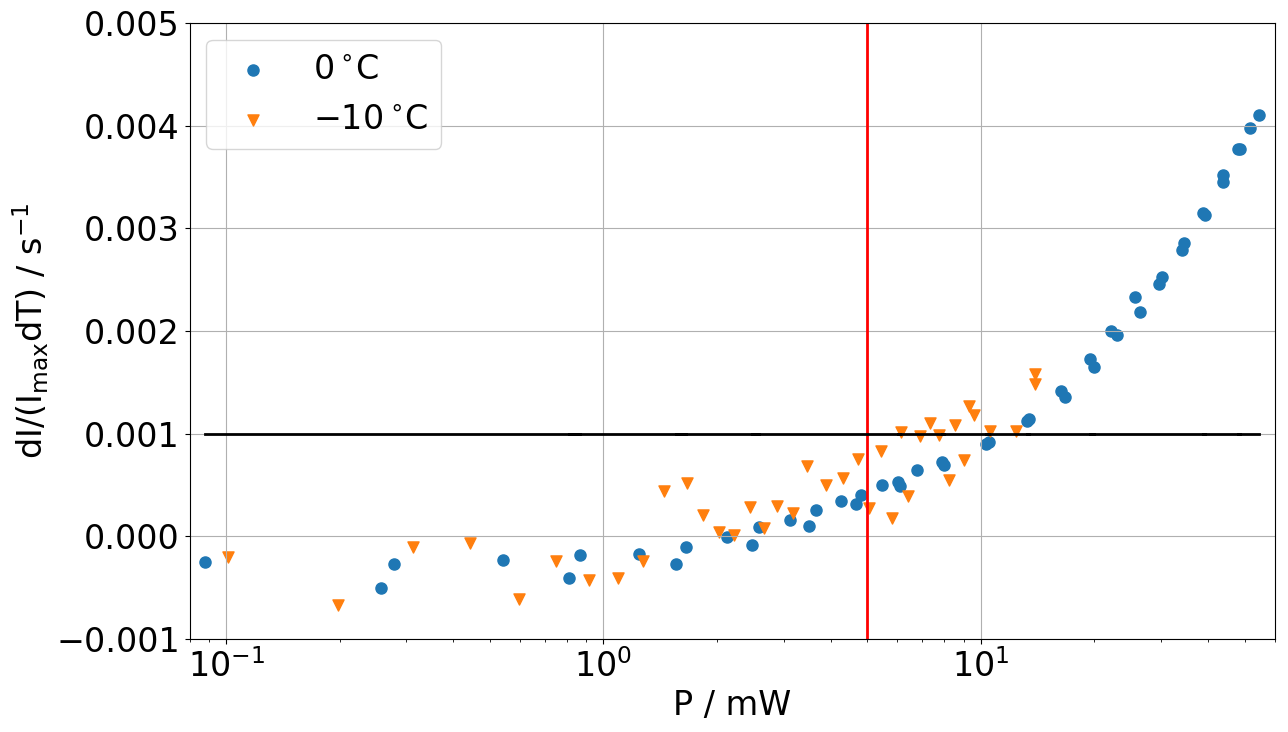}
        \caption{\label{fig:cut} Slope of the leakage current in the first $30\,$s after setting the bias voltage from $0\,$V to a target voltage normalized to the maximal current at that voltage as function of the diode power. The power limit at $5\,$mW and an increase of $1 \text{\textperthousand} / \text{s}$ are highlighted.}    
\end{figure}

A source of uncertainty for the measurement of the sample temperature is the self-heating of the diode. 
A potential small temperature increase can not be measured using the thermistor due to its distance from the diode.
Therefore, measurements are excluded where self-heating leads to significant deviations between measured and actual diode temperature.
To identify these measurements, a power limit is determined experimentally by measuring the leakage current over time after setting the voltage from $0\,$V to a target bias voltage at the highest possible slew rate and monitoring the current for $30\,$min. 
This is done for bias voltages up to $1000\,$V at temperatures of $0\celsius$ and $-10\celsius$. 
If self-heating is present, the leakage current is expected to keep increasing after the bias voltage has settled.
If this is not observed, the available cooling power is assumed to be sufficient to suppress this effect and operate the diode at a stable temperature.

In figure \ref{fig:cut}, the slope of a linear fit to the leakage current data in the first $30\,$s after applying the target voltage is shown as a function of the maximal diode power observed in the measurement. 
To be able to compare the slope at different voltages and temperatures, the slope of the leakage current, normalized to the maximum during the measurement, is plotted.
A slope above $1 \text{\textperthousand} / \text{s}$ is interpreted as an indicator of significant self-heating, resulting in a power limit of $5\,$mW.

\section{Results}
\label{sec:results}
\subsection{Determining \eeff as a function of electric field}
\label{sec:eeff}
\eeff is determined following the methodology presented in ref. \cite{Reiner}. 
In this method, the equation 
\begin{equation}
    \label{eqn:step0}
    I(T) = A T^2\exp\left(- \frac{\eeffnospace}{2 k T}\right)\text{,} 
\end{equation}
with the proportionality factor $A$ specific to the device, its size and the bias voltage is used. It is rearranged to isolate the unknown parameters and the logarithm of this equation is calculated resulting in  
\begin{equation}
    \label{eqn:step1}
   \ln\left(\frac{I(T)}{T^2}\right) = \ln(A) - \frac{\eeffnospace}{2 k T}\text{,} 
\end{equation}
which can be rearranged into
\begin{equation}
    \label{eqn:step2}
   - 2 k T \ln\left(\frac{I(T)}{T^2}\right) = \eeffnospace - 2 k T \ln(A) \text{.} 
\end{equation}
A linear function is fitted to the left side of \ref{eqn:step2}:
\begin{equation}       
    \label{eqn:step3}
    \text{p1(T)}  = - 2 k T \ln\left(\frac{I(T)}{T^2}\right)\text{.}
\end{equation}
Comparison between \ref{eqn:step2} and \ref{eqn:step3} shows that the y-axis intercept of p1(T) corresponds to \eeff while the slope can be used to determine the proportionality factor $A$.

Using this method, \eeff can be determined for each bias voltage, excluding measurements where the above mentioned power limit is reached.

\begin{figure}[ht]
    \centering
    \includegraphics[width=.7\textwidth]{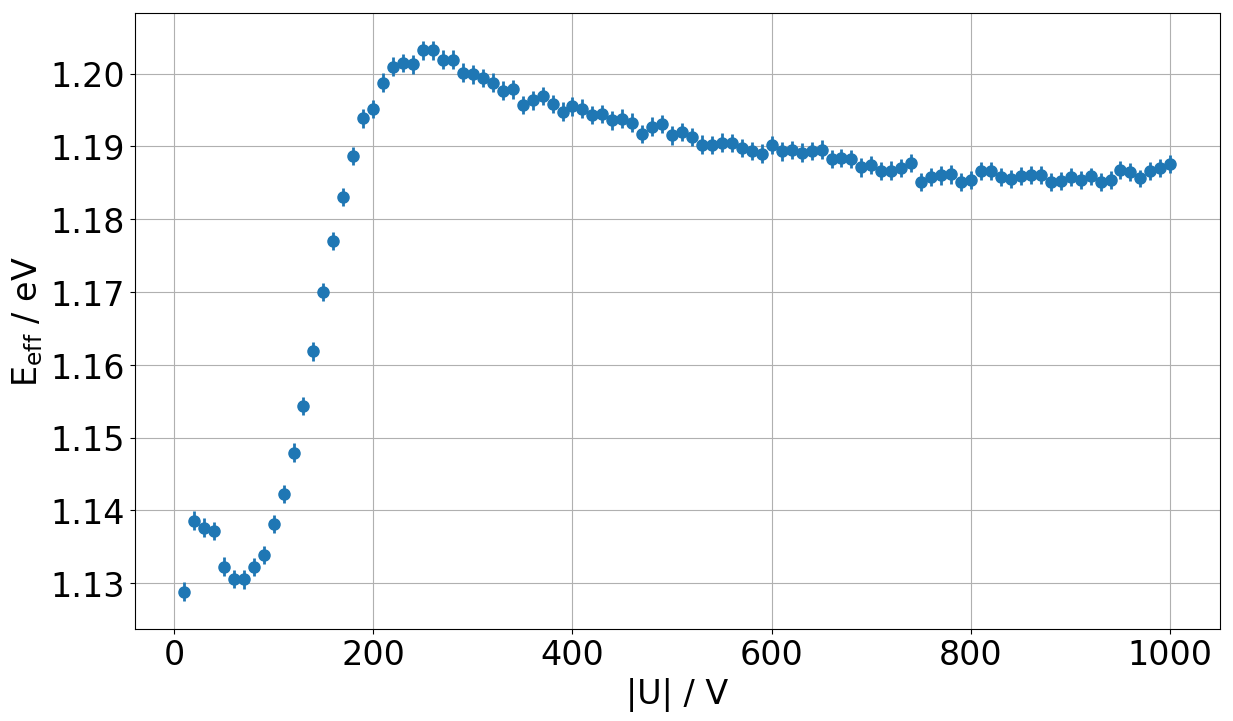}
    \caption{\label{fig:curve} \eeff versus bias voltage for sample P3 before annealing in the temperature range $-30\celsius$ to $-20\celsius$.}
\end{figure}

A clear dependence between the applied bias voltage (and therefore the average electric field in the bulk) and \eeff is observed for all investigated diodes. 
Figure \ref{fig:curve} shows the dependence for sample P3: At low bias voltages the value for \eeff is around $1.13\,$eV, rising steeply to a maximum value around $1.20\,$eV, and decreasing slightly for higher bias voltages.
In case of the sample P3, the maximum is reached at $250\,$V which is compatible with the full depletion voltage estimated from the onset of the plateau region in the I-V characteristic (see figure \ref{fig:bulkcmp}).
A similar behaviour is observed for samples P1 and P5, which is shown in figure \ref{fig:fluence39}.
Sample P4 does not reach depletion voltage before the power limit is exceeded.

\subsection{Influence of the temperature interval}
\label{sec:temp}
\begin{figure}[ht]
    \centering
     \includegraphics[width=.7\textwidth]{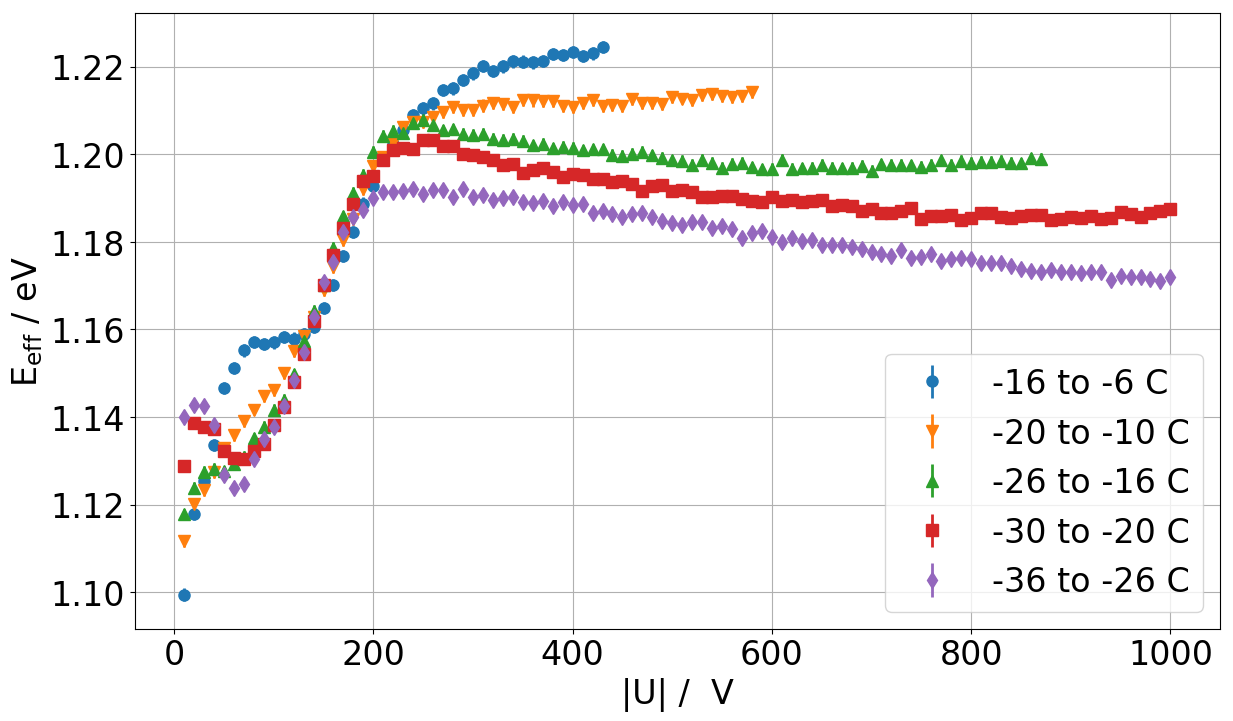}
    \caption{\label{fig:P3_Tdep} \eeff is shown against the applied bias voltage for P3 before annealing for different temperature intervals.}
\end{figure}

A dependence of \eeff on the temperature range of the I-V measurements was observed during this investigation.
To quantify this effect, \eeff is determined for different temperature intervals.
An interval width of $10\,$K was chosen to include sufficient data points while keeping the interval small.
In figure \ref{fig:P3_Tdep}, the bias-voltage dependence of \eeff for multiple of these intervals is plotted for the diode P3 before annealing.
Significant differences between the curves for different temperature intervals at voltages above the depletion voltage are visible where lower temperatures lead to a lower \eeff.

This effect is visible for all investigated samples and a similar effect was observed in ref. \cite{Chili13}.

\begin{figure}[ht]
    \begin{subfigure}[t]{0.49\textwidth}
    \centering
    \includegraphics[width=\textwidth]{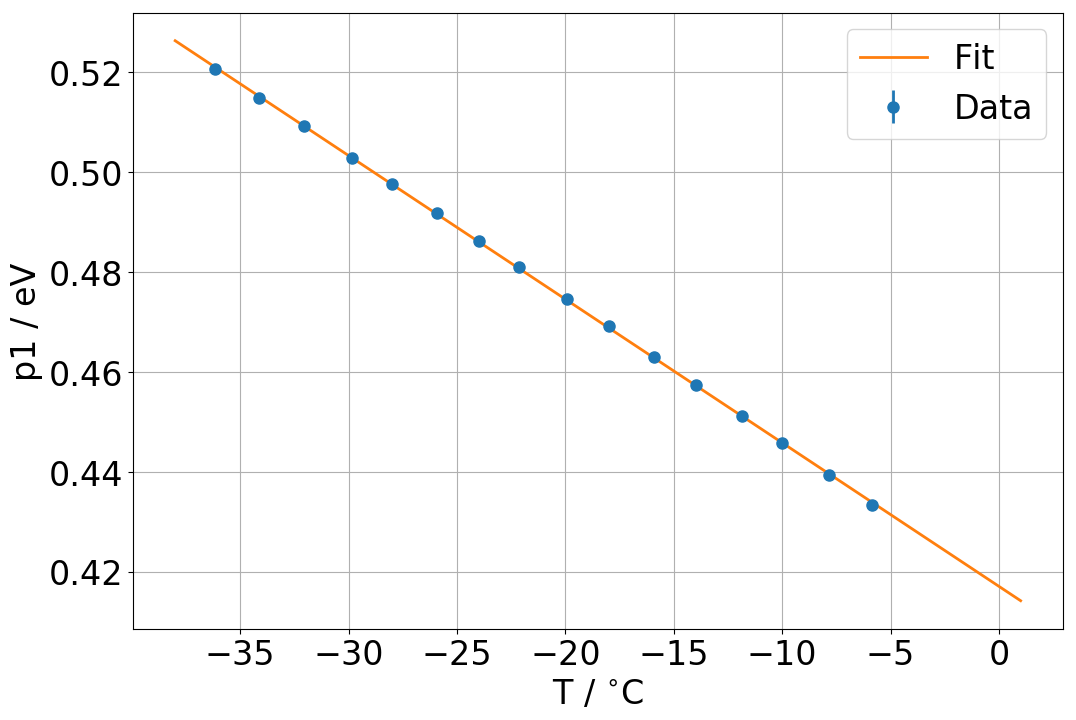}
    \caption{\label{fig:fit}}
    \end{subfigure}
    \hfill
    \begin{subfigure}[t]{0.49\textwidth}
    \centering
    \includegraphics[width=\textwidth]{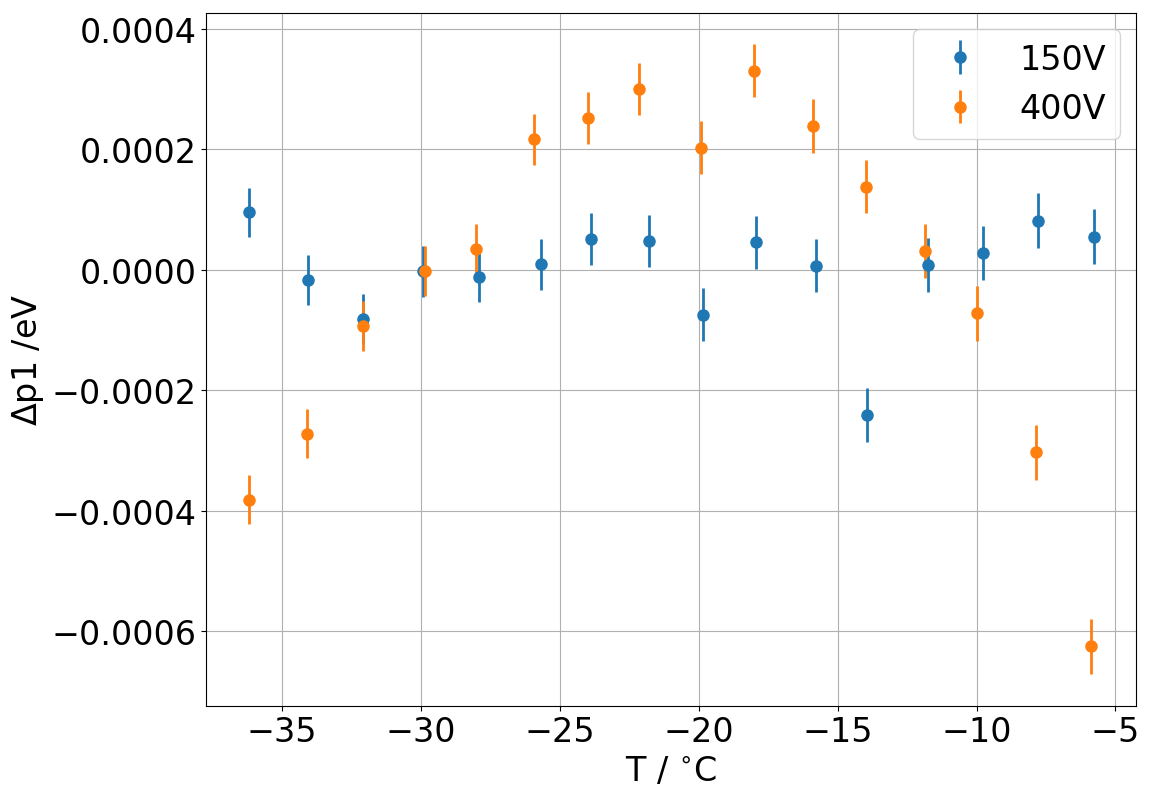}
    \caption{\label{fig:P3_dp1}}
    \end{subfigure}
    \caption{a) The fit of p1(T) is shown for diode P3 at $400\,$V for temperatures from $-36\celsius$ to $-6\celsius$. b) The difference between measured and fitted p1(T) at $150\,$V and $400\,$V is shown for P3 before annealing.}
\end{figure}

In figure \ref{fig:fit}, the linear fit to determine \eeff is shown for temperatures between $-36\celsius$ and $-6\celsius$ at $400\,$V. The residuals of this fit are shown in figure \ref{fig:P3_dp1} in comparison to the residuals of the fit at $150\,$V. At $150\,$V, the residuals are small and randomly distributed around 0 while at $400\,$V, the larger residuals show a clear temperature dependence. 

Self-heating can be excluded as cause for this dependence because it would result in a correlation between power dissipation and the onset of this effect. However, the onset appears at similar voltages for all temperature intervals and is not correlated with the power dissipation. 

Further investigations use the interval from $-30\celsius$ to $-20\celsius$ to minimise the influence of this effect.

\subsection{Comparison of samples at different fluences and annealing stages}
After irradiation, additional defects in the silicon lead to an increased leakage current. It is investigated if this influences the temperature scaling behaviour of the leakage current.

\begin{figure}[h]
    \begin{subfigure}{0.49\textwidth}
    \centering
    \includegraphics[width=\textwidth]{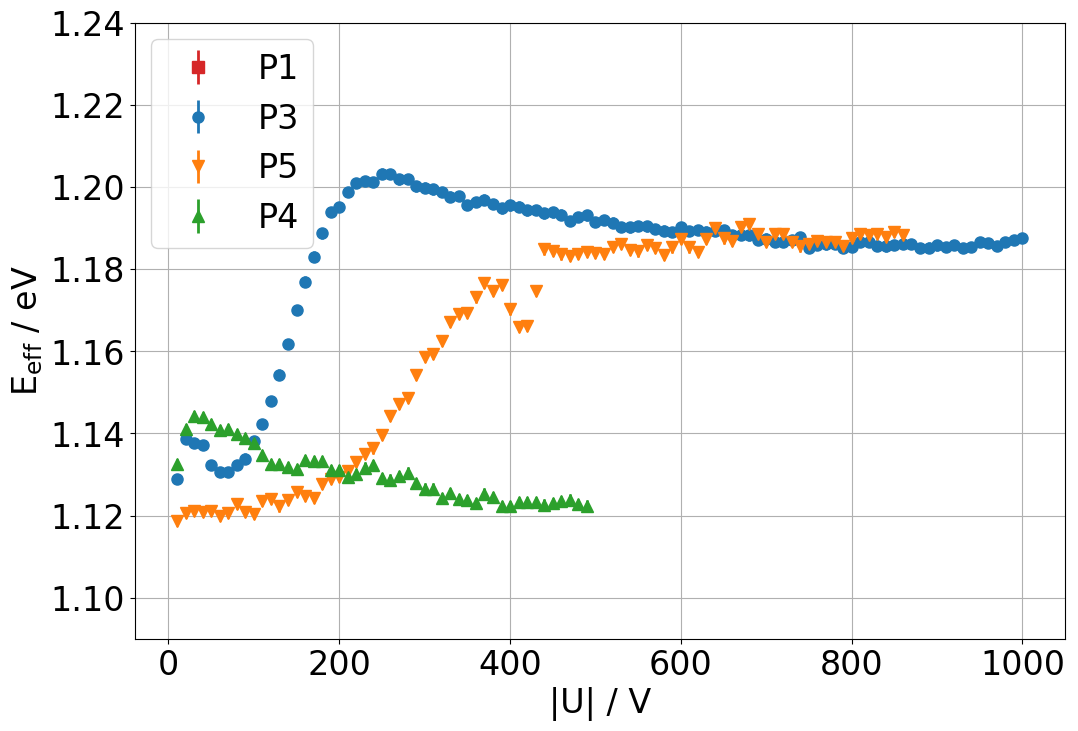}
    \caption{\label{fig:fluence0}}
    \end{subfigure}
    \hfill
    \begin{subfigure}{0.49\textwidth}
    \centering
    \includegraphics[width=.95\textwidth]{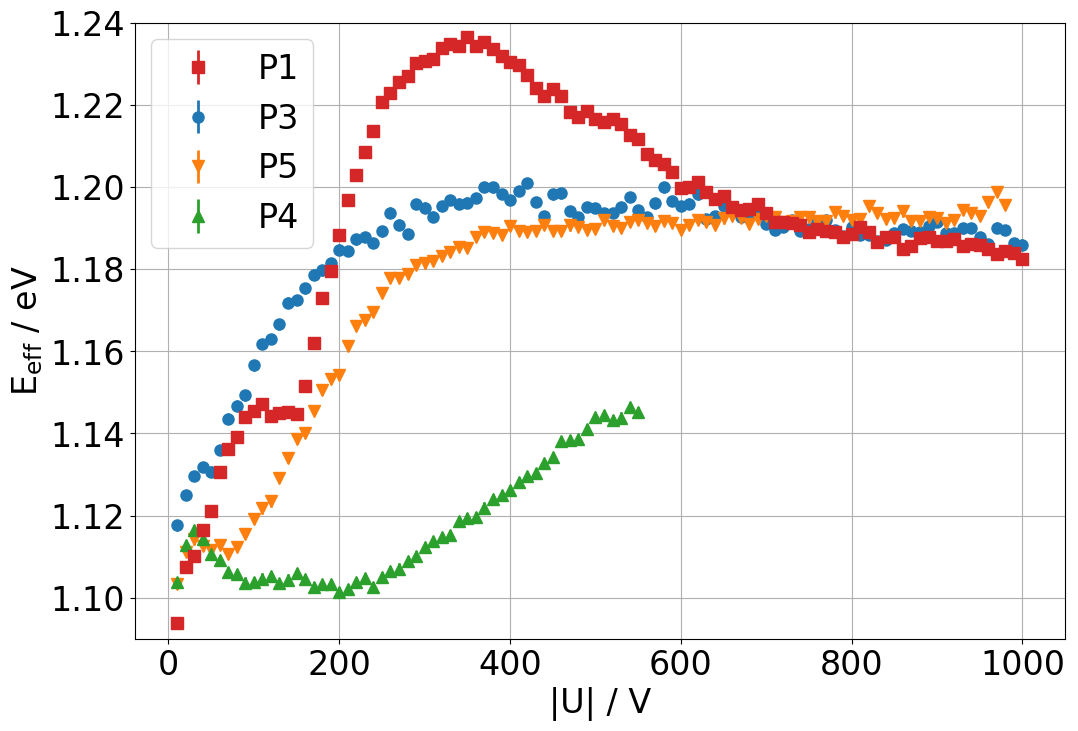}
    \caption{\label{fig:fluence39}}
    \end{subfigure}
    \caption{\label{fig:fluences} a) \eeff as a function of the bias voltage before annealing for samples irradiated with fluences from $7 \times 10^{14}\,\neqpcm$ to $3 \times 10^{15}\,\neqpcm$ for the temperature range $-20\celsius$ to $-30\celsius$. b) The same is shown after $1170\,$min of annealing at $60\celsius$.}
\end{figure}

\begin{figure}[h!]
    \begin{subfigure}{0.49\textwidth}
    \centering
    \includegraphics[width=\textwidth]{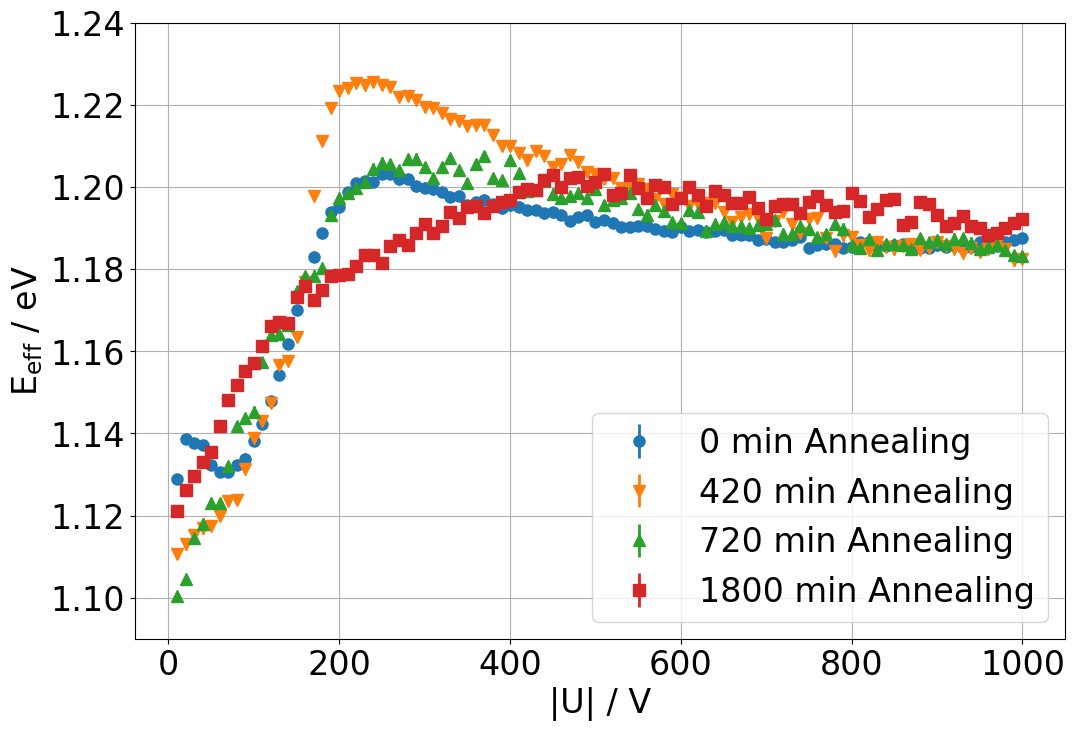}
    \caption{\label{fig:annP3}}
    \end{subfigure}
    \hfill
    \begin{subfigure}{0.49\textwidth}
    \centering
    \includegraphics[width=\textwidth]{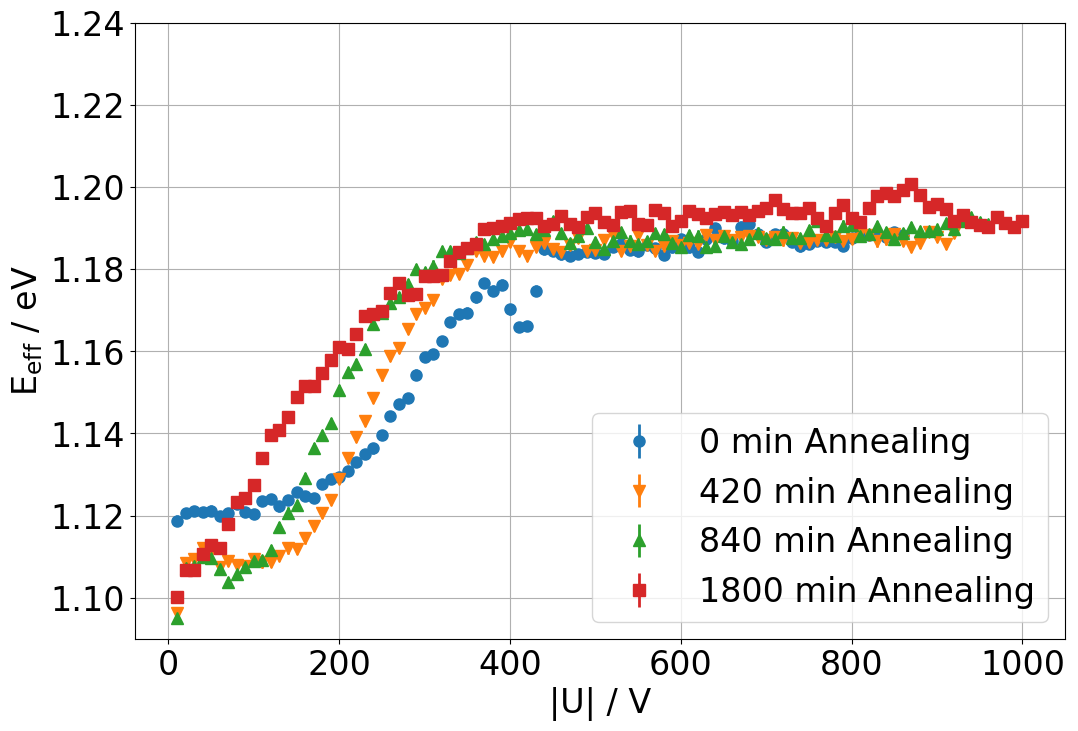}
    \caption{\label{fig:annP5}}
    \end{subfigure}
    \caption{\label{fig:ann} \eeff versus bias voltage during annealing for a) sample P3, b) sample P5.}
\end{figure}

Figure \ref{fig:fluences} shows \eeff for all samples (a) before annealing and (b) after annealing for $1170\,$min at $60\celsius$.
Before annealing, P3 ($7 \times 10^{14}\neqpcm$) reaches its maximal \eeff at about $250\,$V.
For P5 with a higher fluence of $1 \times 10^{15}\neqpcm$, the maximal \eeff is shifted to $400\,$V.
For P4 ($3 \times 10^{15}\neqpcm$), \eeff declines from around $1.14\,$eV at $0\,$V to $1.12\,$eV at $500\,$V. 
After annealing, the characteristic rise of \eeff is shifted to lower bias voltages for all samples.  
A detailed study of this shift is shown in figure \ref{fig:ann}, where the \eeff curves of P3 and P5 are plotted for intermediate annealing steps.

For P3, the value of \eeff measured at 250V increases with annealing up to $420\,$min and decreases afterwards.
In the last measurement after $1800\,$min of annealing, \eeff at $250\,$V is below its value before annealing.
The diode P5, with a fluence of $1 \times 10^{15} \neqpcm$, seems to show a qualitatively different behaviour than P3:
The increasing slope, and therefore the onset of the plateau at $1.19\,$eV of \eeffnospace, shifts to lower voltages with annealing.

The differences in annealing behaviour between P3 and P5 as well as the difference between the samples P1 and P3 in \ref{fig:fluence39} are possibly due to the annealing history during irradiation. However, other effects cannot be ruled out and further measurements are necessary to investigate this.

\section{Discussion}
\label{sec:disc}
This study observed a dependence of the temperature scaling factor \eeff on the electric field. The investigated samples show values between $1.10\,$eV and $1.14\,$eV at the lowest voltages which increase towards a maximum of $1.19\,$eV to $1.23\,$eV.
For samples irradiated to fluences up to $1\times10^{15}\neqpcm$, \eeff reaches a plateau compatible with $1.19\pm0.01\,$eV at high bias voltages. 
The voltage needed to reach this plateau is compatible with the respective depletion voltage estimated from the IV curves which were used to determine \eeffnospace.
The sample with a higher fluence of $3\times10^{15}\neqpcm$ does not reach this plateau before exceeding the power limit.

The development of this plateau with fluence explains the lower values of \eeff for highly irradiated samples observed in other studies.
This can be seen in ref. \cite{Wiehe}, where \eeff was observed to decrease down to $1.14\,$eV in a sample with a fluence of $2\times 10^{16}\,\neqpcm$. This value was determined by averaging $\eeff$ from measurements between $0\,$V and $1000\,$V. This sample can be assumed to be not fully depleted at $1000\,$V.
In addition to the dependence on the applied electric field, changes with annealing were observed in the region of the estimated full depletion voltage.

A hypothesis presented in ref. \cite{Chili13} is that an active electrically neutral bulk (ENB) contributes to the leakage current: charges generated in the ENB are pulled into the space charge region by an electric field in the ENB. The electric field in the ENB has been observed in charge collection efficiency measurements \cite{CCE} as well as Edge-TCT-measurements \cite{TCT}. 
The amount of charge carried into the space charge region is then hypothesised to have a temperature dependence resulting in a decreased \eeff  if an active ENB is present. 
This hypothesis is compatible with results from this study.

Further investigations with more samples and additional fluences would lead to a better understanding of this effect.
This would allow selecting better values of \eeff for scaling the leakage current with temperature for irradiated detectors without determining it specifically for each device.

\section{Summary}
\label{sec:summary}
This study investigates the temperature scaling of leakage current generated in the bulk of proton irradiated silicon diodes with fluences up to $3 \times 10^{15}\,\neqpcm$.

The scaling parameter \eeff was determined as a function of the applied electric field for applied voltages from $0\,$V to $1000\,$V.
The measured values of \eeff for voltages above full depletion voltage are $1.19\pm0.01\,$eV, slightly lower than the value of $1.21\,$eV \cite{Chili13} measured in previous studies.

This study shows a dependence of \eeff on the applied electric field.
For the investigated samples, \eeff is measured to be as low as $1.10\,$eV at low voltages and increasing until full depletion is reached. 
This behaviour is affected by annealing.

\acknowledgments

The work presented here is carried out within the framework of Forschungsschwerpunkt FSP103 and supported by
the Bundesministerium f\"ur Bildung und Forschung BMBF under Grants 05H15PECAA and 05H15PECA9.

\bibliography{ms}
\bibliographystyle{JHEP}

\end{document}